\documentclass[12pt]{iopart}

\usepackage{iopams}
\usepackage{graphicx}
\usepackage[normal]{caption}
\usepackage{color}
\usepackage{float}
\usepackage{changepage}

\begin{document}

\title[Towards the explicit computation of Bohm velocity]{Towards the explicit computation of Bohm velocities associated to N-electron wave-functions with arbitrary spin-orientations}

\begin{center}
\author{A. Alarc\'{o}n, X.Cartoix\`{a}, and X.Oriols}
\end{center}

\address{Departament d\rq{}Enginyeria Electr\`{o}nica, Universitat Aut\`{o}noma de Barcelona, 08193, Bellaterra, SPAIN}
\ead{xavier.oriols@uab.es}

\begin{adjustwidth}{-1cm}{0.5cm} 
\begin{abstract}
  The direct solution of the many-particle Schr\"{o}dinger equation is computationally inaccessible for more than very few electrons. In order to surpass this limitation, one of the authors [\textit{X. Oriols, Phys. Rev. Lett. 2007, 98 (066803)}]  has recently proposed a new model to study electron-electron correlations from Bohm trajectories associated to time-dependent wave-packets solutions of pseudo single-particle Schr\"{o}dinger equations. In the aforementioned paper only the orbital exchange interaction is considered assuming that all electrons have the same spin orientation. Then, the many-particle wave function is a complex Slater determinant of the single-particle wave-packets. In the present work the previous formalism is extended to study many-particle wave functions where the electrons have different spin orientations.The main difficulty to treat \textit{N} different electron spin orientations  with time-dependent wave-packets is that one must study all the possible \textit {N!N!} products of permutations  among spin states. To overcome this computationally inaccessible problem, in this article the total wave function is treated as a separated product of two many-particle wave functions, the first with spin up and the second with spin down. In order to numerically justify this approximation, the Bohm velocity in different antisymmetric total wave-function scenarios is computed. The computational results confirms the accurate validity of our approximation under a large number of cases.
\end{abstract}
\end{adjustwidth}

\pagebreak

\section{Introduction}

From a computational point of view, the direct solution of the many-particle Schr\"{o}dinger equation is inaccessible for more than very few electrons.
This computational problem is at the heart of almost all the unsolved problems in quantum mechanics.
In the field of quantum electron transport in nanoscale devices, which is the topic and the motivation of this work, the standard procedure to overcome this computational
barrier is to assume noninteracting (Fermi liquid) electrons and to decouple the study of transport from that of the electronic structure (via the effective electron mass)\cite{rlandauer70PM,sdatta00SM,wrfrensley90RMP,mbuttiker90PRL}. Nowadays, ab initio quantum transport approaches, based on the density functional theory (DFT)\cite{wkohn65PR}, are being developed \cite{mbrandbyge02PRB,skurth06PRB}, to surpass the previous approximations.

Recently, Bohmian mechanics has undergone a revival in order to develop new quantum computational algorithms. For example, one of the authors has recently published a novel formalism \cite{xoriols07PRL} to model quantum electron transport with Coulomb and exchange correlations using Bohm trajectories. These Bohm trajectories are associated to a wave-packet solution of a pseudo single-particle Schr\"{o}dinger equation with time-dependent potential profiles.
In fact, the time dependence of the potential profile associated to each electron cannot be neglected because it includes the correlations with all other electrons.
Therefore, such single-particle wave-packets have in principle no orthogonal orbitals. This introduces an additional computational difficulty when dealing with the exchange interaction.

In reference \cite{xoriols07PRL} only the orbital exchange interaction is considered. In others words, the formalism in \cite{xoriols07PRL} guarantees that Bohm trajectories in the configuration space are identical when initial electron positions are interchanged. Thus, since the formalism guarantees that the orbital part of the wave-function is antisymmetric, the spin part will necessarily be symmetric.
In general, there is no computational problem to antisymmetrize a wave function when all electrons have the same spin direction. It can be performed by
means of Slater determinants. The problem arises when we try to antisymmetrize wave functions when the particles have different spin orientations.
In this case, it is mandatory to take into account, in the construction of the antisymmetrical wave function, all the permutations products among the different spin states.
Therefore, for \textit{N} electrons, the explicit evaluation of \textit {N!N!} products of permutations is intractable for more than very few electrons (note that $8!^{2}=40320^{2}$).

In the present article, we propose an approximation to the explicit computation of Bohm velocities associated to \textit{N}-electron wave-functions with arbitrarily spin orientations.
For this purpose, we assume that the many-particle wave-function can be separated into a product of spin-up ($\uparrow$)and spin-down($\downarrow$) many-particle wave functions:

\begin{adjustwidth}{-1.7cm}{0cm} 
\begin{eqnarray}
 \Psi\left({\bf \textit{r}_{1}, \textit{r}_{2}, \textit{r}_{3}}...\uparrow_{1},\downarrow_{2},\downarrow_{3}...,\right)\approx
 \phi_{\uparrow}\left({\bf \textit{r}_{1}, \textit{r}_{4}}...\uparrow_{1},\uparrow_{4}...,\right)\cdot\phi_{\downarrow}\left({\bf \textit{r}_{2},\textit{r}_{3}}
...\downarrow_{2},\downarrow_{3}...,\right)
\label{alfonso:ec1}
\end{eqnarray}
\end{adjustwidth}
where we use an uncoupled spin-basis $\{|\uparrow_{i}\rangle\  |\downarrow_{i}\rangle\}$, which is adequate
for (non-conservative spin) open systems.

Using the right hand side of expression (1), the numerical difficulties in the computation of the many-particle
wave function disappear because it can be computed again from a complex
matrix (Slater) determinant. A formal demonstration of the previous approximation well be published somewhere else.
Here, we well provide qualitative arguments about the validity of expression (\ref{alfonso:ec1}) and we will provide numerical examples testing its accuracy.

Finally, it is important to remark that the main effort in solving the many-particle Schr\"{o}dinger equation is done in quantum chemistry, which in general deals with very few particles in closed systems, and later extrapolated to other fields. For example, DFT is now moving from the calculation of ground-state energies towards the computation of electron current in open systems.  In this sense, let us emphasize that the practical computation of an anti-symmetrical wave-function with arbitrary spin orientations has been extensively treated in the quantum chemistry literature.
However, the requirement of a well-defined total system spin is not necessary when studying quantum electron transport because the number of particles in the open system is not fixed and our device is not rotationally invariant.
Therefore, the uncertainty of the total system spin because of computational simplifications (eg. spin contamination) will not be an issue of relevance for this work.

\section{Numerical Results}

We consider a system of \textit {N} electrons described by a many-particle wave-function:

\begin{adjustwidth}{3cm}{0cm} 
\begin{equation}
\Psi\left({\bf \textit{r}_{1},\textit{r}_{2},\textit{r}_{3}}...\uparrow_{1},\downarrow_{2},\downarrow_{3}...\right)
\label{ec2}
\end{equation}
\end{adjustwidth}
where ${\bf \textit{r}_{i}}$  is the electron position and $\uparrow_{i}$/$\downarrow_{i}$
its (up /down ) spin.
Then, we construct the many-particle wave-function  by antisimmetrizing  single-particle states.
Each single-particle state (i.e. orbital) is a gaussian wave-packet with initial central position $X_{0}$, initial wave-vector $K_{0}$ and spatial dispersion $\sigma$. Such initial "orbitals" are quite common in quantum electron transport modeling and they tend to a scattering state for large spatial dispersions. For simplicity, in this preliminary work, we consider "neutral" electrons (i.e. without Coulomb interaction among them or other particles).

The Bohm velocity of each electron $i$ has to be
computed directly from the many-particle wave-function as follows:

\begin{equation}
v_{i}\left({\bf \textit{r}_{1},\textit{r}_{2},\textit{r}_{3}}...\uparrow_{1},\downarrow_{2},\downarrow_{3}...\right)=
\frac{\langle J_{i}\left({\bf \textit{r}_{1},\textit{r}_{2},\textit{r}_{3}}...\uparrow_{1},\downarrow_{2},\downarrow_{3}...\right)\rangle}
{|\Psi\left({\bf \textit{r}_{1},\textit{r}_{2},\textit{r}_{3}}...\uparrow_{1},\downarrow_{2},\downarrow_{3}...\right)|^2}
\label{ec3}
\end{equation}
where the quantum current for a particular electron $i$ is
\begin{adjustwidth}{-1cm}{0cm} 
\begin{equation}
 \langle J_{i} \rangle=\frac{\hbar}{m}
 \rm{Im}\left[ \Psi\left({\bf \textit{r}_{1},\textit{r}_{2},\textit{r}_{3}}...\uparrow_{1},\downarrow_{2},\downarrow_{3}...\right)\times
 \frac{\partial\Psi\left({\bf \textit{r}_{1},\textit{r}_{2},\textit{r}_{3}}...\uparrow_{1},\downarrow_{2},\downarrow_{3}...\right)}
{\partial x_{i}} \right]
\label{ec4}
\end{equation}
\end{adjustwidth}
and $|\Psi\left({\bf \textit{r}_{1},\textit{r}_{2},\textit{r}_{3}}...\uparrow_{1},\downarrow_{2},\downarrow_{3}...,\right)|^2$ is the total norm of the many - particle wave-function.

In order to numerically verify the correctness of equation (\ref{alfonso:ec1}), we compute
the Bohm velocity associated to electron 1 in three different
situations keeping the antisymmetry of the total wave-function: First, when the electron 1 is alone (See figure 1). Then, we compute the velocity of electron 1 when it is surrounded by four additional exchange-interacting electrons.  We use the expression "exchange-interacting electrons" in reference only at the interaction among electrons as results of Pauli's exclusion principle. In Fig. 2, we use the left hand side of expression (1), while in Fig. 3 we use the right hand side (i.e. the approximation proposed in this work).

\begin{figure}[h]%
\centering
\includegraphics*[width=0.7\linewidth]{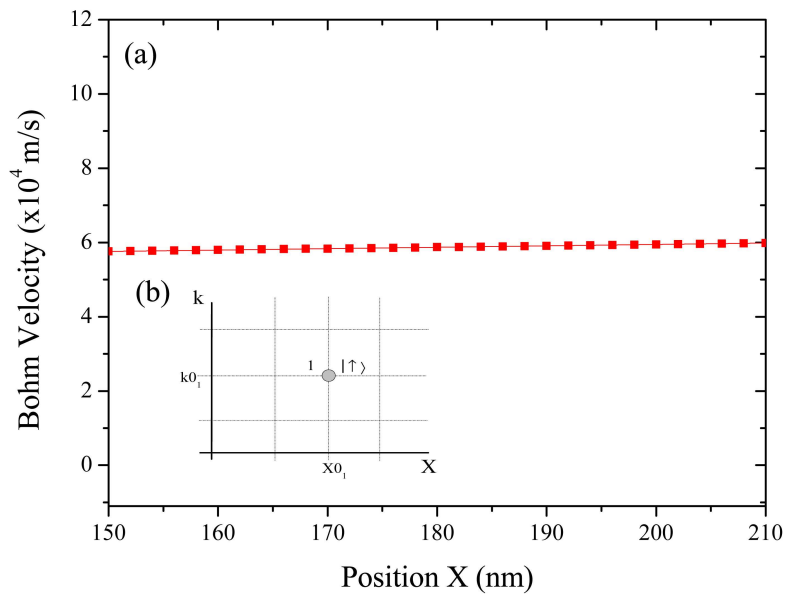}
\caption{%
  {\small (a) Bohm velocity for an independent electron. (b) Schematic representation of the system for an electron where we indicate the central value of the $X_{0}$ and wave-vector $K_{0}$ of initial wave-packet.}}
\label{Fig1}
\end{figure}

In detail, we show in Figure~\ref{Fig1} the computation of the Bohm velocity (with an approximate value of 6 x $10^{4}$
m/s) for one independent (spin-up) electron along different positions.

\begin{figure}[h]%
\centering
\includegraphics*[width=0.7\linewidth]{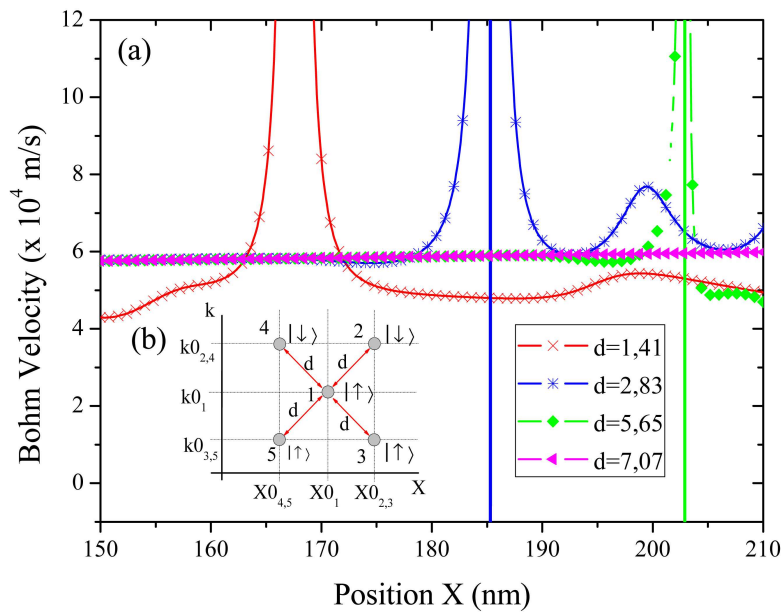}
\caption{%
  {\small (a) Bohm velocities for 1-electron using different values of \textit{d} for a system of 5 electrons (3 spin-up and 2 spin-down). (b) In this scheme we indicate the central value of the $X_{0}$ and wave-vector $K_{0}$ of initial wave-packet.}}
\label{Fig2}
\end{figure}

\begin{figure}[h]%
\centering
\includegraphics*[width=0.7\linewidth]{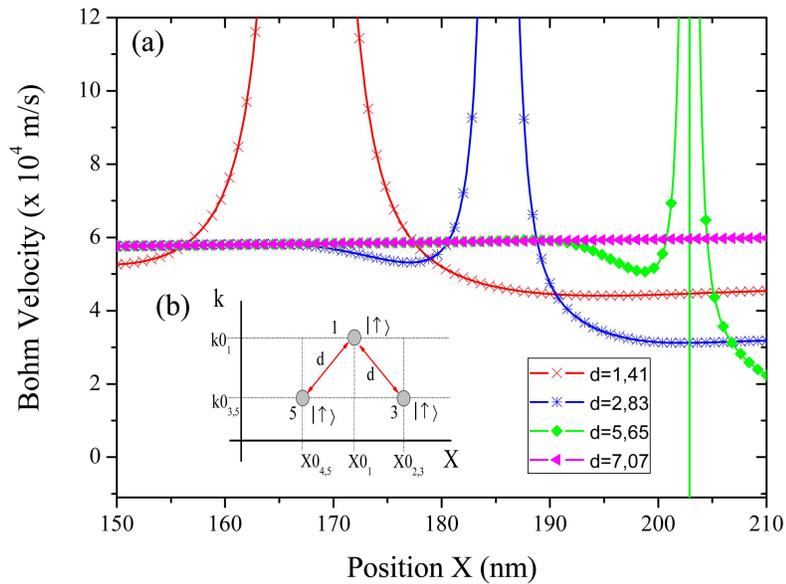}
\caption{%
  {\small (a) Bohm velocities for 1-electron using different values of \textit{d} for a system of 3 electrons (spin-up). (b) In this scheme we indicate the central value of the $X_{0}$ and wave-vector $K_{0}$ of initial wave-packet. }}
\label{Fig3}
\end{figure}

In Figure \ref{Fig2},
we plot the exact computation of Bohm velocity for a system of 5 electrons
studying the electron 1 (see inset) when other 4 exchange-interacting electrons are present. We define the parameter \textit{d} as a normalized (i.e. without units) phase-space distance \cite{xoriols04Nano} between electron 1 and the others (see insets in Figs.~\ref{Fig1},~\ref{Fig2},
and ~\ref{Fig3}).
Then, we compute the Bohm velocity for four different values of distance  \textit{d}  among electrons:\textit{1.41,2.83,5.65} and \textit{7.07}.
For large \textit{d} (left triangle line) the velocity of the electron 1 is not changed by the position of the other 4 electrons. But when we decrease  \textit{d} (cross line), the Bohm velocity becomes very different from the result of Fig.~\ref{Fig1} as a consequence of the Pauli (Exclusion) principle.

In Figure~\ref{Fig3}, we consider the same system as in Fig. ~\ref{Fig2}, but we compute the Bohm velocity using the right hand side of Eq.(~\ref{alfonso:ec1}). Thus, when we compute the Bohm velocity (Eq.(~\ref{ec3})) of the particle 1  we observe that only the many - particle wave function with spin up $\phi_{\uparrow}\left({\bf \textit{r}_{1},\textit{r}_{4}}...\uparrow_{1},\uparrow_{4}...,\right)$ will contribute the computation of Bohm velocity.

The strong resemblance
between the Bohm velocities of Figs.~\ref{Fig2} and~\ref{Fig3}
from different values of \textit{d} provides a numerical justification of the approximation mentioned in Eq.(\ref{alfonso:ec1}) for the computation of many-particle Bohm velocities.
Similar results are obtained for many other spin schemes.

\begin{figure}[H]%
\centering
\includegraphics*[width=0.7\linewidth]{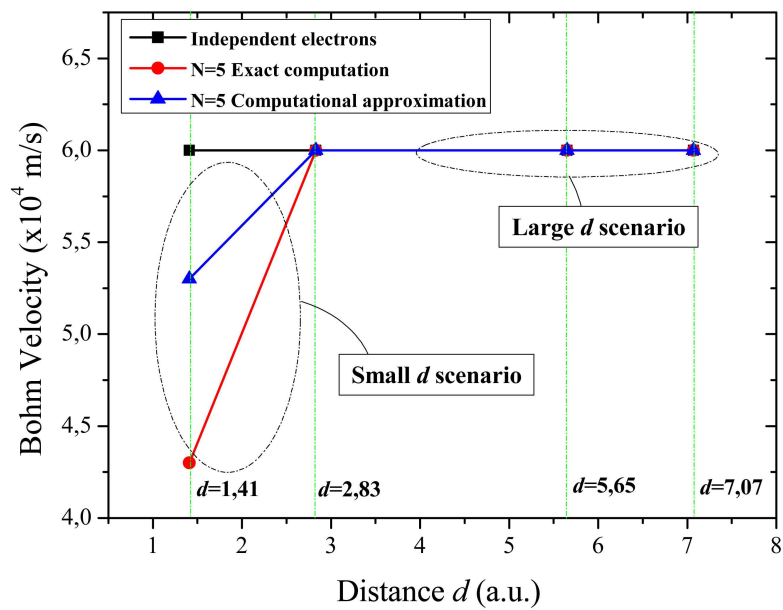}
\caption{%
    {\small For a particular position ($X_{0}=150nm$) of Bohm velocity of Figs~\ref{Fig2} and~\ref{Fig3}. We plot the Bohm velocity in function distance \textit{d} among electrons for three different electron scenarios: independent electron, exact computation and computational. Lines are a guide to the eye.}}
\label{Fig4}
\end{figure}

Despite the similitude between the two previous schemes, we find some differences in the plotted Bohm velocities. We explain these differences in Fig.~\ref{Fig4}. We choose a particular position ($X_{0}=150nm$ in Figs.~\ref{Fig2} and~\ref{Fig3}) and we plot the Bohm velocity as a function  of the distance \textit{d} for independent electron (solid black line), the exact computation (dashed red line) and  our computational approximation (dotted blue line). In Figure~\ref{Fig4} we observe two different regions (circle dotted lines): region
with for small \textit{d} and region with large \textit{d}.

In order to better understand the differences that we observe for the small \textit{d} scenario between the exact computation and our computational approximation, it is necessary to treat with detail the total norm.
We compute the exact total norm as a \textit {N!N!} sum of terms which are scalar products between single-particle wave functions.
We can divide the total norm into two parts: principal contribution and spurious contribution. The principal contribution is computed from the left hand of expression (\ref{alfonso:ec1}). The spurious contribution is computed from the product of single particle terms that are present only in the left hand but are not present in the right hand of expression (\ref{alfonso:ec1}).

\begin{figure}[h]%
\centering
\includegraphics*[width=0.7\linewidth]{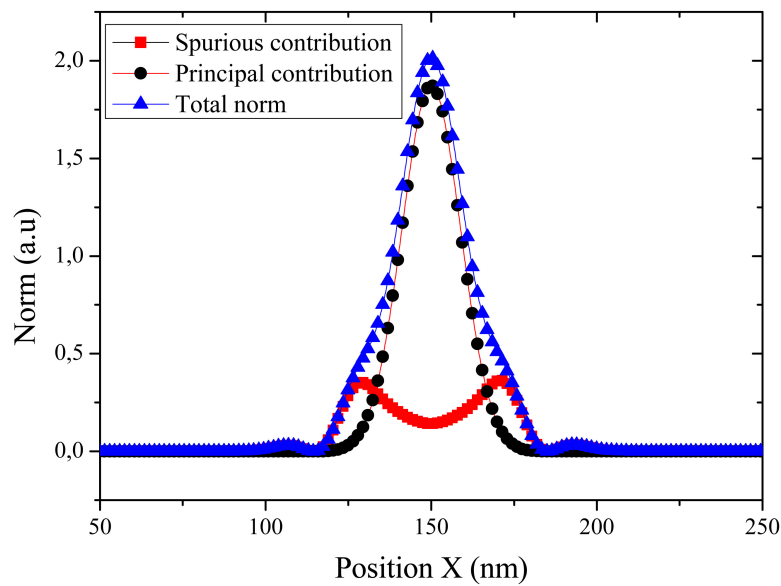}
\caption{%
    {\small Total norm for small \textit{d} scenario. The total norm is divided  by  principal contribution (black line) and spurious contribution (red line). In this case we find a significant contribution of spurious contribution.}}
\label{Fig5}
\end{figure}

\begin{figure}[H]%
\centering
\includegraphics*[width=0.7\linewidth]{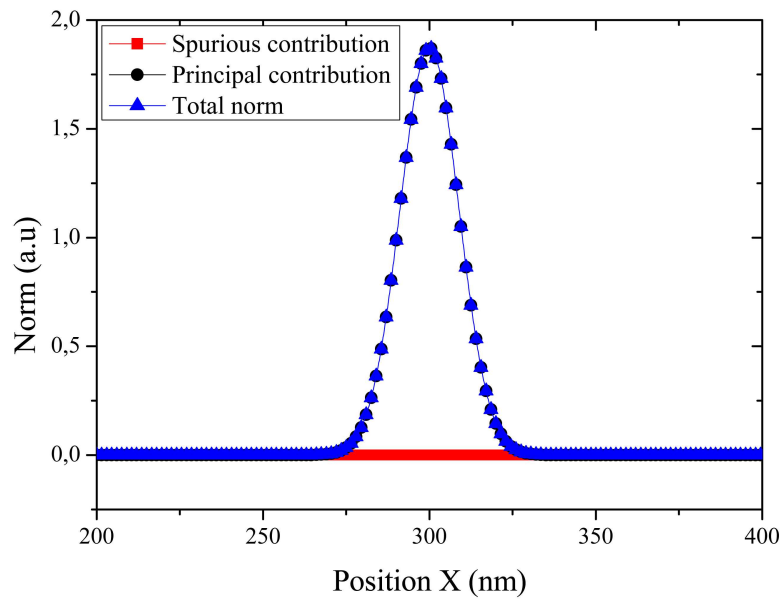}
\caption{%
    {\small Total norm for large \textit{d} scenario. The total norm is divided  by  principal contribution (circle solid  black line) and spurious contribution (square solid red line). In this case the spurious contribution  is almost zero.}}
\label{Fig6}
\end{figure}

In both figures~\ref{Fig5} and~\ref{Fig6} we plot the principal contribution (circle solid  black line) and spurious contribution (square solid red line) in the scenario described in Fig.~\ref{Fig4} for small \textit{d} and large \textit{d} respectively. If we compare these two figures we find that in the small \textit{d} scenario an important role of spurious contribution. Or the contrary, if we observe  the large \textit{d} scenario the spurious contribution is almost zero this is an expected result.

\section{Conclusions}

In this article we have presented an approximate expression, Eq.(\ref{alfonso:ec1}), with the intention to surpass the computational difficulties associated to  many-particle wave-functions with different spin orientations when dealing with quantum electron transport modeling (in open system where total system spin is not well-defined).
In detail, the main difficulty to treat \textit{N} time-dependent wave-packets with different electron spin orientations is that one must study all the possible products of permutations (\textit {N!N!}) among spin states when we antisymmetrize the wave-function.
In this work, we propose that many-particle wave-functions can
be separated into a product of spin up and spin down many-particle wave-functions.
To justify the validity of Eq.(\ref{alfonso:ec1}) we have computed the Bohm velocity in three different situations keeping the total wave-function.
The practical viability of our proposal can be used for studying
systems with a large (N$\simeq$100) number of electrons using complex
Slater determinants.
The approximation of Eq.(~\ref{alfonso:ec1}) is a trade-off between accuracy and computational viability as we show in Fig.~\ref{Fig4} and it has significant implications in quantum electron transport with Coulomb and exchange interactions among electrons.
In the near future, this approach
will be applied for the computation of the average current or its fluctuations
\cite{galbareda08PRB} in zero or high frequency \cite{aalarcon09JSTAT}
quantum scenarios.

\section*{acknowledgement}
This work was supported through Spanish MEC project MICINN TEC2009-06986.



\end{document}